# POSSE: PATTERNS OF SYSTEMS DURING SOFTWARE ENCRYPTION

David A. Noever and Samantha E. Miller Noever

PeopleTec, Inc., Huntsville, Alabama, USA
david.noever@peopletec.com

## ABSTRACT

This research recasts ransomware detection using performance monitoring and statistical machine learning. The work builds a test environment with 41 input variables to label and compare three computing states: idle, encryption and compression. A common goal of this behavioural detector seeks to anticipate and short-circuit the final step of hard-drive locking with encryption and the demand for payment to return the file system to its baseline. Comparing machine learning techniques, linear regression outperforms random forest, decision trees, and support vector machines (SVM). All algorithms classified the 3 possible classes (idle, encryption, and compression) with greater than 91% accuracy.

## 1. INTRODUCTION

Ransomware generally refers to malicious software that encrypts a victim's file system before demanding payment for an unlocking key. The attack combines file cryptography with traditional malware and phishing techniques, but economically derives its profitability from the apparent international anonymity of cryptographic currency markets like Bitcoin. Depending on the software type and origin of the attackers, files may first be exfiltrated, then leaked as proof of the attack, or published as privacy breaches. Even if the victim maintains a strong file backup schedule, the data leakage alone may prove too damaging and subsequently force the victim's payment even without receiving a valid unlocking key. The economics of ransomware-for-hire and crypto-currency anonymity have combined globally to unleash 304.7 million attempted attacks in the first 6 months of 2021 (up 151% annually) and according to FBI tracking, have spawned more than 100 different strains [1].

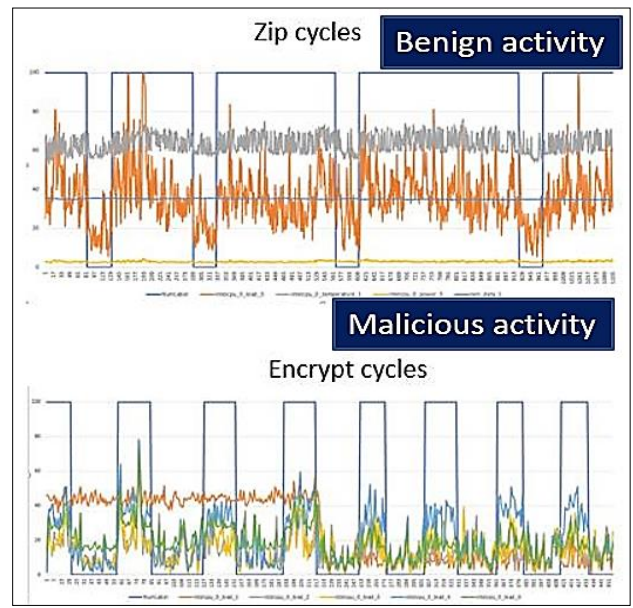

Figure 1. Pattern of Hardware Sensors to Predict Activity

Like traditional phishing attacks, a malicious email attachment file downloads, or social engineering a credential may grant the attacker initial system access. Initial access may quickly infiltrate entire networks (e.g. worms) and cascade the locking to halt a large enterprise in minutes to hours. Privacy breaches have proved lucrative to attackers who target medical infrastructure while operating concerns have motivated payment by other critical sectors such as pipelines. Like malware prevention, traditional methods have been tried to halt the ransomware epidemic, including blocking IP addresses, patching vulnerabilities, and expert business rules for defining normal vs. abnormal activities. The unique challenge however has proven the maintenance of recent backups coupled with an encryption timeline such that restoring older files does not spawn reinfection. Recent efforts [1-17] have explored novel ways to detect the initial early stage of unusual activities using pattern recognition and deep learning methods.

This work addresses novel behavioral indicators during software encryption. The approach applies machine learning (ML) methods for pattern recognition. To examine the feasibility of anomaly detection from secondary hardware changes, we experimentally explore a test harness for cycling different types of software-intensive stages systematically (Figure 1). We alternate busy and idle times through thousands of repetitions. We divide the behavior into three classes based on normal background activities alternating with malicious and benign software steps. The model problem mirrors many of the early stages of a ransomware attack, principally its overworking of the CPU, GPU, and file system during software encryption. Previous work has proposed a sensor-based approach that identifies hardware indicators of early ransomware stages [2-13]. By hardware monitoring during malicious events, previous research has further applied deep learning methods to understand malware infection [14-15], with specialized efforts applied to industrial control systems [16] and sophisticated side-channel attacks such as Spectre and Meltdown [17]. A common goal of this behavioral detector centers on anticipating and short-circuiting the final step of hard-drive locking with encryption and the demand for payment to return the file system to its baseline.

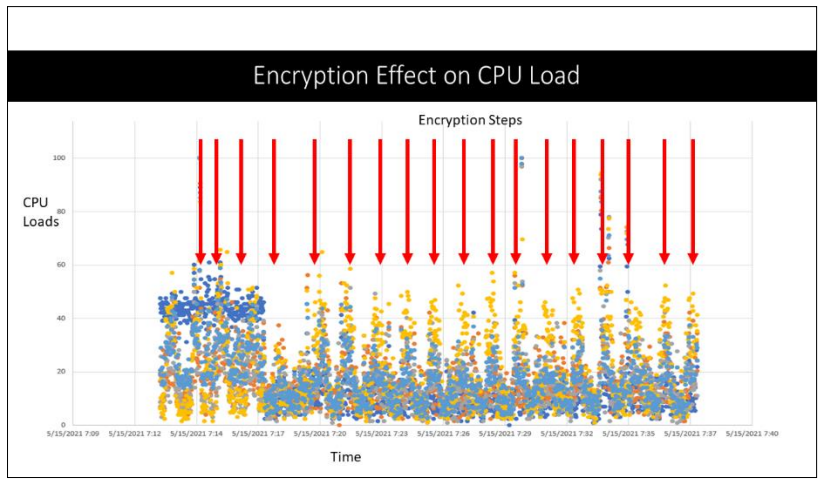

Figure 2. Activity Spikes Alternating Idle and Encryption Steps

## 2. METHODS

This work describes initial experiments using blockchain encryption while monitoring hardware usage spikes (1000 trials). The simplest approach might forensically explore the normal and abnormal extremes for CPU performance using Windows Resource Monitors, such as PerfMon. For a single machine, these logs provide a usage record of CPU, disk, network, and memory. Most forensic or event tracing monitoring however relies on anomaly detection with additional analytic tools like a Security Information Event Manager (SIEM). We apply the Open Hardware Monitor

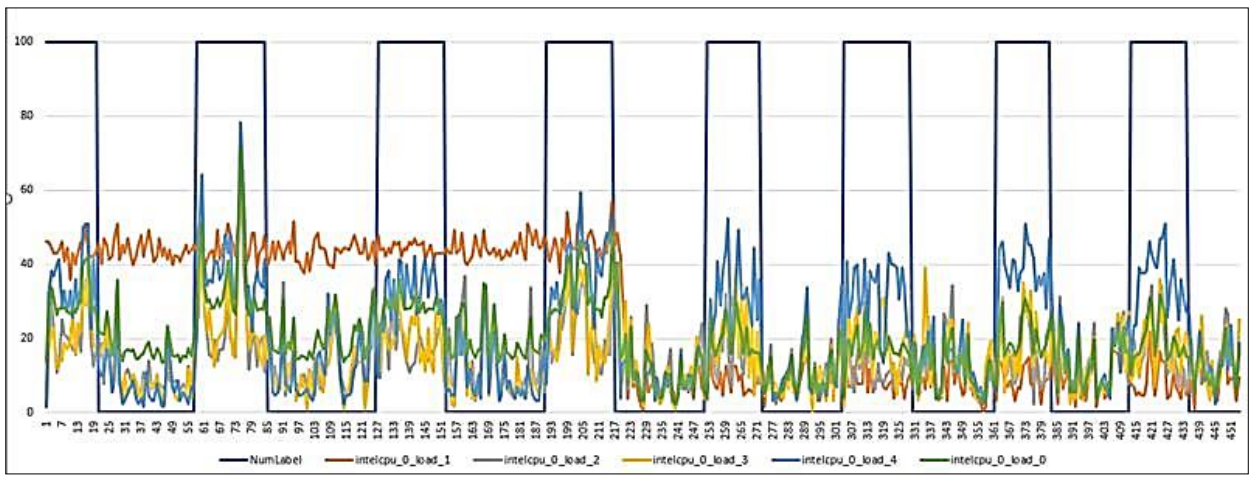

Figure 3. Encryption Cycles and Overlay for Hardware

[18] to log 41 performance metrics and create a controlled time series for a single Windows machine (Figure 3). The x64-based laptop computer specification includes an Intel 4-core processor (i7-7700 HQ CPU @ 2.8 GHz), an NVIDIA Quadro M1200, and a Microsoft Windows 10 operating system. The hardware log (comma-separated values) features an adjustable time step (set to 1-second refresh). The performance metrics include both GPU and CPU load,

temperature, bus clock speed, RAM load (memory, used and available), and dual hard drive load, temperature, and used space. The hypothesis to test in this setup follows from previous work showing that CPU-intensive encryption can be an early indicator of ransomware.

To generate representative training data, we apply neutral (idle) conditions alternating with different software functions. Following the model of RanSim [19], we treat normal file compression (zip) as a benign but CPU-intensive operation and customize an Advanced Encryption Standard Cipher Block Chaining mode (AES-CBC) encryptor [20] as a model malicious software cycle. Previous work [21] has analyzed the AES-CBC time to encrypt in a Crypto-Wall ransomware attack on Internet of Things (IoT) devices. For simulations, we chose a simple block cipher (Blowfish) which can encrypt and decrypt messages of arbitrarily long length. We used a relatively short block length (1024 bytes) and a 56-long hex key ('ffffffff x 7') as the encryption/decryption passphrase. The padding method ("space"), initialization vector IV ('$KJh#(}q'), and randomly generated salt value were chosen for simplicity but the resulting encrypted file is compatible with OpenSSL standards. Ransomware attackers have applied a host of weak (e.g., Vigenere cipher) and strong encryption algorithms [21-23] ranging from AES-256 to RSA-4096.

To generate benign activity in a repeatable way, we applied the deflate algorithm (Zip) as a lossless data compression that uses LZSS and Huffman coding [24]. To maintain contact with the encryption stages, we apply compression without encryption or password usage to the same target (plain-text) files. Because compression took longer than encryption to finish, the spacing intervals between benign and malicious software applications differed. The training data offered a class distribution ratio of 599:1181:5219 for Encryption : Idle: Compression.

For both encryption and compression, the target files ranged by an order of magnitude in size (0.1-1 GB) and consisted of random ASCII text files generated using all alphabetic characters and numerical digits. Each simulation created 150-1000 compressed or encrypted versions of the same file, which was subsequently deleted and the operating system returned to idle background operations for 50 seconds before initiating another step cycle. Figures 2-3 illustrate a typical cycle of activity-rest that spans approximately 20 encryption cycles while monitoring CPU load as its activity indicator. In total, the simulations collected 116,029 intervals (in seconds for 32 hours) of 41 hardware-dependent indicators of activity (4.7 million data points). For data analysis, the labeled set (0=background, 1=encrypt, 2=compress) was randomly shuffled and split into 70-30 ratios applying standard train-test partitions. Using R (caret) and python (sklearn) library packages, we classified whether a given timestamp with the 41 hardware performance parameters is encryption or not.

| Approach | Error | Avg Class Error |
|---|---|---|
| Decision Tree | 4.1 | 8.06 |
| Random Forest | 2.6 | 6.1 |
| SVM | 4.2 | 8.7 |
| Linear | 2.6 | 5.8 |

Figure 4. Error Rates for Statistical ML Algorithms

## 3. RESULTS

Figure 4 shows the error rates for statistical machine learning algorithms. Linear regression outperforms random forest, decision trees, and support vector machines (SVM). All algorithms classified the 3 possible classes (idle, encryption, and compression) with greater than 91% accuracy.

An outcome of applying random forest to behavioral analytics stems from assigning feature or factor importance (Figure 5). Of the 41 hardware variables, the compression class depends strongest on hard drive (2) load. The encryption class depends strongest on CPU power. The broad range of variable importance and multiple factor contributions suggest that the decision boundary separating each class is complex and cannot be rendered from a threshold approach. For example, assigning a certain CPU power threshold for warning the potential ransomware victim appears to be infeasible, or at least generates many false-positive events during normal CPU usage.

| Hardware Parameter | Idle | Encrypt | Compress |
|---|---|---|---|
| hdd_2_load_0 | 13.29 | 10.47 | 41.88 |
| intelcpu_0_load_4 | 14.69 | 12.2 | 22.43 |
| intelcpu_0_power_1 | 21.46 | 26.04 | 19.9 |
| intelcpu_0_power_0 | 17.97 | 20.56 | 18.42 |
| intelcpu_0_temperature_4 | 10.46 | 15.38 | 13.17 |
| intelcpu_0_load_2 | 10.05 | 10.22 | 12.37 |
| intelcpu_0_power_3 | 12.44 | 32.67 | 12.01 |
| intelcpu_0_load_1 | 9.64 | 8.66 | 11.41 |
| ram_data_1 | 8.44 | 10.35 | 11.19 |
| intelcpu_0_load_0 | 10.48 | 13.33 | 11.15 |
| ram_data_0 | 7.79 | 11.65 | 11.12 |
| hdd_0_load_0 | 5.42 | 9.67 | 11.05 |

Figure 5. Variable Importance for the Three State Classes

# 4. CONCLUSIONS AND FUTURE WORK

This work has explored a test harness that monitors hardware performance indicators to identify alternating cycles of benign and malicious events. Two benign classes include idle background noise and a false signal (compression). The malicious class includes a true signal (encryption) that matches well to a known ransomware algorithm (CBC) seen in the wild. All the machine learning algorithms show greater than 90% accuracy to classify the current computing state based on observing the system performance alone. To supplement this classification and reduce false positives, future enhancements can provide traditional indicators of compromise. In its simplest implementation, the working prototype would restrict processes that launch taxing performance hits prior to ransomware completion.

**ACKNOWLEDGMENTS**
The author would like to thank the PeopleTec Technical Fellows program for its encouragement and project assistance.